\documentclass[aps,prl,showpacs,reprint]{revtex4-1}
\usepackage{hyperref}
\usepackage{amsmath}
\usepackage{braket}
\usepackage{natbib}
\usepackage{graphicx} % needed for figure
\usepackage{dcolumn} % needed for some tables
\usepackage{amssymb}
\usepackage{color}

\newcommand{\eqaref}[1]{Eq.~\eqref{#1}} 
\newcommand{\figref}[1]{Fig.~\eqref{#1}}  

\renewcommand{\bf}[1]{\mathbf{#1}}

\begin{document}
\title{Chiral specific electron vortex beam spectroscopy}
\author{J. Yuan}
\author{S. M. Lloyd}
\author{M. Babiker}
\affiliation{Department of Physics, University of York, Heslington, York, YO10 5DD, UK}
\date{\today}

\begin{abstract}

Chiral electron vortex beams, carrying well-defined orbital angular momentum (OAM) about the propagation axis, are potentially useful as probes of magnetic and other chiral materials.  We present an effective operator, expressible in a multipolar form, describing the inelastic processes in which electron vortex beams interact with atoms, including those present in Bose-Einstein condensates, involving exchange of OAM.  We show clearly that the key properties of the processes are dependent on the dynamical state and location of the atoms involved as well as the vortex beam characteristics.  Our results can be used to identify scenarios in which chiral-specific electron vortex spectroscopy can probe magnetic sublevel transitions normally studied using circularly polarized photon beams with the advantage of atomic scale spatial resolution.
\end{abstract}

\maketitle
Particle vortices, most notably electron vortices (EVs), are currently the focus of much interest following the prediction by Bliokh \textit{et al.} \cite{Bliokh2007} and their experimental realization in a number of laboratories, using various techniques \cite{Uchida2010,Verbeeck2010,McMorran2011a,Gnanavel2012, Schattschneider2012b, Verbeeck2012}.  This area has recently emerged after much fruitful research has been carried out on optical vortices (OVs) over the last two decades or so, which has led to a wealth of fundamental knowledge and significant applications \cite{Allen2003, Yao2011a, Andrews2012}.  Both the optical and electron vortices  are characterised by the singular nature of their wavefronts, with a well defined vortex core and quantised orbital angular momentum (OAM) about the vortex axis.  The general expectation is that in all cases the vortex OAM should play an important role in the interaction of the vortex with matter.  However, in the case of an OV, a dipole active transition involves exchange of OAM with the centre of mass only \cite{Babiker2002, Lloyd2012b}, a finding which has been confirmed experimentally \cite{Araoka2005a, Loffler2011}. The development of OAM-based OV beam spectroscopy has been hampered by the weakness of optical multipolar transitions.  In contrast, we have recently demonstrated theoretically that OAM can be transferred efficiently from an EV beam to atomic electrons through dipole active transitions \cite{Lloyd2012, Lloyd2012b} and experimentally a dichroic electron energy-loss spectroscopic signal has been detected \cite{Verbeeck2010}, opening up the prospect of chiral specific electron vortex beam spectroscopy (CEVBS) based on OAM selection rules.  Using a new analytical method, we present an effective operator in the context of OAM transitions in quantum systems using electron vortex beams.  This is important for the realization of CEVBS as it allows the derivation of the key OAM- and chiral-related characteristics, going beyond the derivation of the dipole OAM selection rules to also include a multipolar expansion and the spatial dependence of the quantum transitions involved.  The new results suggest that a confocal spectroscopy set-up could be used to obtain optical activity or X-ray circular dichroic spectroscopy at atomic resolution, for characterization of chiral or magnetic materials and for the determination of the coherent state of a cold-atom condensate.

The leading interaction between the EV and an atom possessing $Z$ electrons is given by the Coulomb interaction Hamiltonian
\begin{equation}
\hat{H}_{\text{int}}=-\frac{Ze^2}{4\pi\epsilon_0|\bf{r}_v-\bf{R}|}+\sum_{j=1}^Z \frac{e^2}{4\pi\epsilon_0|\bf{r}_v-\bf{r}_j|},
\label{Interacting_Hamiltonian}
\end{equation}
where $\bf{r}_v$, $\bf{r}_j$ and $\bf{R}$ are the position vectors, respectively, of the beam electron, the $j$-th atomic electron and the nucleus, all expressed relative to the laboratory frame of reference.  The transition matrix element between states of the combined atom-vortex system can be written as 
$M_{fi}=\Braket{F|\hat H_{\text{int}}|I}$, where $\Ket{I}$ and $\Ket{F}$ are, respectively, the initial and final unperturbed quantum states of the overall system, being products of unperturbed quantum states of the EV and those of the atom: $\Ket{\psi_\text{{EV}}}\Ket{\psi_\text{{atom}}}$.  In the present case the atomic quantum state can be taken as a product of the quantum state of its nucleus, here taken to also be characterized by the centre of mass of the atom, and that describing the internal electronic state relative to the centre of mass  $\bf{R}$, i.e. $\Ket{\psi_\text{{atom}}}=\Ket{\psi_{cm}(\bf{R})}\Ket{\psi_{q}(\bf{r}_1,...\bf{r}_j,...\bf{r}_Z)}$.

Without loss of generality, we assume a Bessel EV beam of winding number $l$  which is an eigenstate of the Schr\"{o}dinger equation for a free electron in cylindrical polar coordinates $(\rho_v,\phi_v,z_v)$ with the beam axis along the $z$-direction,
\begin{equation}
\Ket{\psi_\text{{EV}}}=\Ket{k_{\perp}, l, k_z}_{\text{lab}}=\frac{\sqrt{k_{\perp}}}{2\pi}J_l(k_{\perp}\rho_v)e^{il\phi_v+ik_zz_v+i\omega t},
\label{ElectronVortexWavefunction}
\end{equation}
where $k_{\perp}$ and $k_z$ are the transverse and longitudinal components of the wavevector of the vortex beam such that  $k_{\perp}^2+k_z^2=k^2=\frac{2mE}{\hbar^2}$, with $E$ the beam energy, and $J_l(k_{\perp}\rho_v)$ is the $l$th order Bessel function.

CEVBS is concerned with processes in which an incident EV mode $\Ket{k_{\perp},l,k_z}_{\text{lab}}$ is scattered by the atom into an outgoing EV  mode $\Ket{k'_{\perp},l',k_z'}_{\text{lab}}$, with the atom undergoing a quantum transition between its internal eigenstates. The treatment can be readily extended to more general vortex beams since such beams can be represented by a linear combination of the Bessel basis modes discussed here.  As a simplification, we shall initially assume that the scattering process does not alter the state of the atomic centre of mass.

In analogy with light interacting with the atom, the transition matrix element for an EV interacting with the atom may be reduced to the following form \cite{Thole1992}
\begin{equation}
\mathcal{M}_{fi}=\frac{e^2}{4\pi\epsilon_0}\sum_j\Braket{f|\hat{O}^{l,l'}_j|i},
\label{MattrixElement_electron}
\end{equation}
where $\Ket{i}$ and $\Ket{f}$ are the initial and final states of the atom.   For convenience, we define an $F$-function as:
\begin{equation}
F_{\alpha}^{m,n}(k,k')=J_m(k_{\perp}\rho_{\alpha})J_{n}(k'_{\perp}\rho_{\alpha})e^{i(m-n)\phi_{\alpha}}.
\label{Eqn_OverlappingIntegral}
\end{equation}
where $m$ and $n$ are integers and $\alpha$ specifies the in-plane vector concerned in terms of its coordinates, $\rho_{\alpha}$ and $\phi_{\alpha}$.
The effective operator $\hat{O}^{l,l'}_j$, acting on a single electron, then emerges in the form
\begin{equation}
\hat{O}_j^{l,l'}=\frac{\sqrt{k_{\perp}k'_{\perp}}}{4\pi^2}\int_{-\infty}^{\infty} \frac{F_v^{l,l'}e^{i(k_z-k_z')z_v}}{\left|\bf{r}_v-\bf{r}_j\right|} d^3r_v,
\label{Effective_Operator}
\end{equation}
Note that the first term in \eqaref{Interacting_Hamiltonian} does not contribute to the matrix element by virtue of the orthogonality of the initial and
final atomic states $\Ket{i}$ and $\Ket{f}$.

The chief difficulty in the evaluation of the effective operator for the vortex beam-atom interaction in \eqaref{Effective_Operator} stems from the fact that the vortex state function is conveniently expressed in terms of the laboratory frame, while the internal atomic states are customarily expressed in spherical coordinates in a frame of reference centred on the atomic centre-of-mass of coordinate $\bf{R}$.  In order to overcome this difficulty, the addition theorem of Bessel functions \cite{Abramowitz1964} can be utilised to represent the original EV beam of mode $l$ as a sum of other vortex states relative to a shifted frame of reference centred on the atomic centre of mass coordinate $\bf{R}$.  The addition theorem reads
\begin{equation}
J_\mu(a)=e^{-i \mu\theta}\sum_{\nu=-\infty}^{\infty}J_{\mu+\nu}(b)J_{\nu}(c)e^{i\nu\varphi},
\end{equation}  
where $a$, $b$ and $c$ are three sides of a triangle, and $\theta$ and $\varphi$ the internal angles between sides $a$ and $b$, and $b$ and $c$, respectively.  Applying this to the triangle formed by the position vectors of the vortex, nucleus and atomic electron, we identify $\bf{r}_c(\rho_c,\phi_c, z_c)$ as the position vector describing the vortex electron relative to the centre of mass, and after some further algebraic manipulation, we find
\begin{equation}
J_l(k_{\perp}\rho_v)= e^{-i l\phi_v}
\sum_{p=-\infty}^{\infty}  J_{l-p}(k_{\perp}\rho_R) J_{p}(k_{\perp}\rho_c) e^{i (l-p)\phi_R}e^{ip\phi_c}.
\label{Eqn_BesselFnExpansion}
\end{equation}
(168)
%The EV state may now be written as the sum of vortex modes relative to the centre of mass
%\begin{equation}
%\Ket{k_{\perp}, l, k_z}_{\text{lab}}=\sum_{p=-\infty}^{\infty}C_{l-p}\Ket{k_{\perp}, p, k_z}_{\text{cm}},
%\label{Eqn_shifted_beams}
%\end{equation}
%where the subscripts refer to the reference frame relative to which the quantum states are described, and the coefficient $C_{l-p}=J_{l-p}(k_{\perp}\rho_R)e^{i(l-p)\phi_R}$.

As expected, the above expansion indicates that the only vortex mode relative to the centre of mass present for an atom located on the beam axis is that for which $p=l$ (as only $J_0(\rho_R=0)\ne0$). However, for an atom not situated on the beam axis, the strength of the atom-centred vortex modes with $p=l+1$ and $p=l-1$ also become significant when the atom is positioned at radial distances of the order of a fraction of  $\tfrac{\alpha_{l,1}}{k_{\perp}}\approx0.1\text{ nm}$, where $\alpha_{l,1}$ is the first zero of the $l$th-order Bessel function, i.e.~within the first ring of the vortex beam.  Thus the immediate consequence of the shift of the axis is the importance of vortex modes of winding numbers different from $l$, relative to the atomic centre of mass frame \cite{Alexandrescu2005a, Schattschneider2013}.  This mode broadening effect, well known in OV research, is a manifestation of the extrinsic property of the orbital angular momentum of vortex beams \cite{ONeil2002}.

Applying the shifted wavefunctions of Eq.\eqref{Eqn_BesselFnExpansion}, the effective operator $\hat{O}^{l,l'}$ (the subscript $j$ will henceforth be dropped) relative to the atomic frame can be written as:
\begin{equation}
\hat{O}^{l,l'}=\frac{\sqrt{k_{\perp}k'_{\perp}}e^{-i(k_z-k_z')z_R}}{2\pi}\sum_{p,p'=-\infty}^{\infty}F_R^{l-p,l'-p'}I_c^{p,p'},
\label{InteractionHamiltonian4EV}
\end{equation}
where
\begin{equation}
I_c^{p,p'}=\int \frac{F_c^{p,p'}e^{i(k_z-k_z')z_c}}{\left|\bf{r}_q-\bf{r}_c\right|} d^3r_c,
\label{Eqn_Beam-electronFactor4b}
\end{equation}
with $\bf{r}_q(\rho_q,\theta_q,\phi_q)=\bf{r}_j-\bf{R}$ and $\bf{r}_c(\rho_c,\theta_c,\phi_c)=\bf{r}_v-\bf{R}$, being the internal electronic and vortex beam coordinates, respectively, both about the atomic centre.
We have also isolated the centre-of-mass factor $F_R$, relative to the atomic frame, from the integral $I_c$ relevant to coupling with the atomic electronic states.

In order to express the matrix element in terms of multipolar contributions, the effective operator needs to be expanded in powers of $\bf{r}_q$.  This can be achieved by invoking the addition theorem for Bessel functions again in order to achieve a separation of the dependence on the atomic electron position variable ($\bf{r}_q$) from that of the EV ($\bf{r}_c$). This is conveniently done by introducing a relative position vector $\bf{s}=\bf{r}_c-\bf{r}_q$.  After some algebraic manipulation, the integral can be written as follows
\begin{equation}
I_c^{p,p'}=\sum^{\infty}_{u,u'=-\infty}F_q^{p-u, p'-u'}I_s^{u,u'},
\label{Eqn_expansion_of_Ib}
\end{equation}
where $I_s^{u,u'}$ is given by
\begin{equation}
I_s^{u,u'}=\int d^3r_s \frac{J_u(k_{\perp}\rho_s) J_{u'}(k'_{\perp}\rho_s)e^{i(u-u')\phi_s}e^{i(k_z-k_z')z_s}}{(\rho_s^2+z_s^2)^{1/2}}.
\end{equation}
The integral $I_s^{u,u}$ can be evaluated by expressing each Bessel function as a coherent superposition of plane waves with phase angle dependent on the topological charge \cite{Abramowitz1964}.  The linear momentum transfer wavevector is defined as $\bf{Q}(\beta)=\bf{k}_f-\bf{k}_i$, where  $\bf{k}_i$ is the wavevector of the plane-wave components of the incident Bessel beam, and $\bf{k}_f$ is that of the outgoing Bessel beam, with $\beta = \phi-\phi'$ the relative azimuthal angle between $\bf{k}_{i}$ and $\bf{k}_{f}$. The integral over the vortex beam spatial variables may now be evaluated as the Fourier transform of the Coulomb potential, leading to the condition $u=u'$:
\begin{equation}
I_s^u=\frac{1}{\sqrt{2\pi^3}}\int_0^{2\pi}\frac{e^{iu\beta}}{Q^2(\beta)}d\beta.
\label{FTQint1}
\end{equation}
We note that $1/Q^2(\beta)$ is the familiar kinetic factor arising in Coulomb scattering and is the chief reason for the importance of dipole active transitions in electron-atom interaction.

The result for the effective operator of the vortex electron beam can now be determined by combining Equations  \eqref{InteractionHamiltonian4EV}, \eqref{Eqn_expansion_of_Ib} and \eqref{FTQint1}: 
\begin{equation}
\hat{O}^{l,l'}=\hat{O}^{z}\frac{\sqrt{k_{\perp}k'_{\perp}}}{4\pi^2}\sum_{p,p'=-\infty}^{\infty}\sum_{u=-\infty}^{\infty} F_R^{l-p,l'-p'}F_q^{p-u,p'-u}I_s^u,
\label{Eqn_effetive_operator}
\end{equation}  
where $\hat{O}^{z}=e^{i(k_z-k'_z)(z_R+z_q)}$ is the effective operator for out-of-plane excitations.  \eqaref{Eqn_effetive_operator} allows a clear description of the effect of the EV beam expansion - contained in the $F_R$ factors - and its implications for the OAM transfer between the EV and the atomic electron - contained in the $F_q$ factors.  We illustrate this by considering the implications for chiral specific vortex electron beam spectroscopy.

\begin{figure}[htpb]
\includegraphics[width=1\columnwidth]{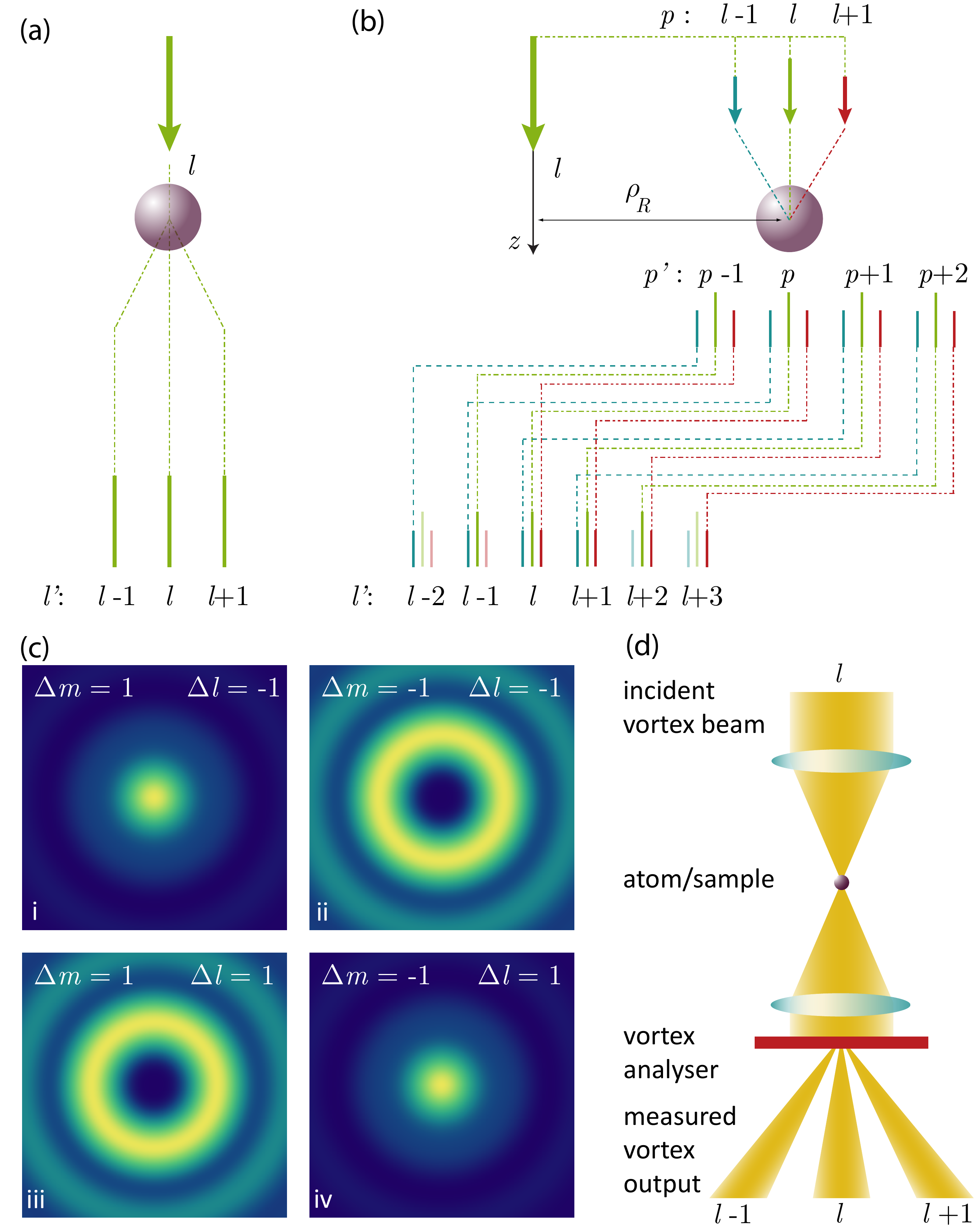}
\caption{(Colour Online) Exchange of OAM in an atom-vortex interaction in the cases when  the atom is situated (a) on the beam axis, and (b) off axis.  In (a) the resulting $l'$ states relate directly to an interaction in which $l-l'$ units of OAM are exchanged; however it is not possible to determine, for example, whether an interaction for which $l-l'=1$ is due to a dipole interaction, or a higher multipole.  In (b), the final states $l'$ arise due to a set of transitions in which varying quantities of OAM are exchanged with the different $p$-modes the atom ``sees". (c) The spatial distribution of inelastic scattering signals for the incident beam with $l=0$ and the outgoing beams with $l'=\pm1$, induced by dipole excitation of magnetic sublevels corresponding $\Delta m=\pm1$.  The images are calculated assuming a 200keV electron beam and the size of the image is $0.4nm\times 0.4nm$, and $k_{\perp}=0.22 {nm}^{-1}$ and $k'_{\perp}=0.11 nm^{-1}$; (d) A confocal arrangement of vortex beams allowing localization of chiral signals from atoms located near the beam axis.}
\label{FIG1}%
\end{figure}

Since the effective operator $\hat{O}^{l,l'}$ acts on the electronic states only through terms containing components of $\bf{r}_q$, only the terms involving $\hat{O}^{z}$ and $F_q$ are relevant. It is clear from the form of $\hat{O}^{z}$ this factor has no chirality feature.  On the other hand, the term $F_q^{p-u,p'-u}$ depends on the in-plane components of  $\bf{r}_q$, and contains the phase factor $e^{i(p-p')\phi_q}$ which is important for chiral specific spectroscopy.  This becomes clear if we consider the simplest case of an atom located on the beam axis, in which case $\rho_{R}=0$.  We then see that $F_R$ is non-zero only for $p=l$ and $p'=l'$, so that the  summation over $p,p'$ in  \eqaref{Eqn_effetive_operator} amounts only to a single term with $p=l$ and $p'=l'$, and $F_{R}=1$.  Using the series expansion of the Bessel functions \cite{Abramowitz1964}, the simplified operator can then be written, in ascending powers of $\rho_q$, 
\begin{align}
\hat{O}^{l,l'}=&\frac{\sqrt{k_{\perp}k'_{\perp}}}{2\pi}e^{-i(k_z-k'_z)z_q}\times\Big[I^{l}_{s}\delta_{l,l'} +\notag\\
&\,\,\left(\mathcal{A}^{+1}e^{i\phi_q}\delta_{l,l'+1}+\mathcal{A}^{-1}e^{-i\phi_q}\delta_{l,l'-1}\right)\rho_q+\mathcal{O}(\rho_q^2)\Big]
\label{Eqn_effective_operator_on-axis}
\end{align}
with $\mathcal{A}^{\pm1} = \frac{1}{2}\left(\pm k_{\perp} I^{l\mp1}_{s} \mp k'_{\perp} I^{l}_{s}\right)$.  Focusing on the dipole-active atomic transition is equivalent to restricting CEVBS to the limit in which the transverse wavevector of the beam is small compared to the inverse size of the systems investigated, a condition often observed in high energy electron energy-loss spectroscopy of atoms \cite{Verbeeck2011}.  In such cases, the dipole terms in \eqaref{Eqn_effective_operator_on-axis}, containing the factors $\rho_q e^{\pm i\phi_q}\delta_{l,l'\pm1}$ operating on the atomic state, causes the magnetic quantum number $m$ of the electronic state to change by one as a result of the transfer of one unit of OAM from or to the EV beam, leading to the dipole selection rule $l-l'=-m+m'=\pm1$.  It is reasuuring that this is precisely the result obtained by Lloyd \textit{et al.} \cite{Lloyd2012, Lloyd2012b} using a completely different approach.  The analysis can be extended to higher powers of $\rho_q$, leading to higher multipole excitations and the associateed selection rules.  This situation is depicted in \figref{FIG1}(a).  For $l=0$, we have $\mathcal{A}^{+1}=-\mathcal{A}^{-1^*}$.  Thus, besides the phase factor, the effective dichroic operator for a vortex beam interacting with an atom is directly comparable to the operator associated with the absorption and emission of either a right ($+$) or left ($-$) handed photon, $\hat{O}^{\pm}\sim (\hat{\pmb{\epsilon}}_x \pm i\hat{\pmb{\epsilon}}_y) \cdot \bf{r}_q=x_q \pm iy_q=\rho_qe^{\pm i\phi_q}$ \cite{Lloyd2012b}.  Because of this formal equivalence, our result is then applicable to any quantum system.  In this regard, CEVBS is similar to EMCD \cite{Schattschneider2006} but would be much more practical because only small angle (i.e. small $Q$) scattering is required in the vortex beam case (\eqaref{FTQint1}), so the signal-to-noise ratio should be much improved.  The case of CEVBS with an atom located near the beam axis is achievable, for example,  with a confocal microscopic arrangement \cite{Frigo2002} adapted for OAM filtered imaging \figref{FIG1}(d), with atomic scale imaging formed by scanning the sample relative to the beam axis.

We can now address the physical meaning of the double summation over $p$ and $p'$ in \eqaref{Eqn_effetive_operator} and the implications of this for the chiral-specific spectroscopy of atoms located away from the vortex beam axis.   The off-axis case is illustrated in \figref{FIG1}(b).  It can be seen that the features uncovered above as regards OAM transfer from the EV to the atomic electron still apply locally at the atom sites, except that now the vortex states with which the off-axis atom interacts are characterised by the winding number $p$, not $l$, and the outgoing states after the electronic transition within the atom are characterised by  $p'$ rather than $l'$. This is the mode broadening effect of the incoming and outgoing EV beams, as described by  \eqaref{Eqn_BesselFnExpansion}.  Since CEVBS is normally conducted with respect to the beam axis, summed over atoms at various off-axis positions, the spectral changes observed in different OAM components of the outgoing EV beam in general can not be exactly related to the change in OAM of the atomic electronic system, as has been assumed in the case of Ref.\cite{Verbeeck2010}.

However, the radial profiles of the incident and outgoing vortex beams can still be chosen to allow individual multipole excitations in the atoms to be probed in such a general case, aided with the knowledge of the effective operator given in \eqaref{Eqn_effetive_operator}.  To illustrate this point, we have taken the simplest case of exciting an atomic magnetic sublevel dipole transition using an incident Bessel electron beam of $l=0$ and examining the probability of finding the outgoing Bessel beams with $l=\pm1$.  The results are shown in \figref{FIG1}(c). For clarity of detail the images have been individually intensity normalised. $\Delta m= \pm1$ refer to the change in the magnetic quantum number induced in the atom.  The probabilities of inducing `allowed' dipole transitions ($\Delta m=-\Delta l=\pm1$) are seen to be strongly peaked at the atom center (Fig. 1c(i, iv)), which is located at the center of the images, with the leading contribution of the order of $|J_0(k_{\perp}\rho_R)J_0(k'_{\perp}\rho_R)|^2$.  The probabilities of inducing `forbidden' dipole transitions ($\Delta m=\Delta l=\pm1$) are, as expected, only significant for off-axis atoms (Fig. 1c(ii,iii), with a much reduced amplitude of the order of $|J_0(k_{\perp}\rho_R)J_2(k'_{\perp}\rho_R)|^2$ and a peak intensity about 3\% of the `allowed' intensity in our simulation.  The dominance of the `allowed' dipole transition for on-axis atoms is due to the narrow radial extent of the incoming and outgoing co-axial Bessel beams involved, a scenario that can be approximated experimentally by the confocal arrangement described in \figref{FIG1}(d).  More importantly, the `allowed' and `forbidden' dipole transitions can be further discriminated as their relative excitation probabilities can be adjusted by varying $k_{\perp}$ and $k_{\perp}'$. Higher multipole atomic transition can also contribute to the outgoing vortex beams, as indicated in \figref{FIG1}(b), but with reduced intensities because of \eqaref{FTQint1}.  The complementarity of the images from the $l'=1$ (Fig. 1c(i,ii) and $l'=-1$ (Fig.c(iii,iv)) channels is just an extension of the complementarity of $A^{+1}$ and $A^{-1}$ factors in \eqaref{Eqn_effective_operator_on-axis} mentioned above.

Detailed knowledge of the effective operator from which the matrix element can be derived can also be used to interpret the general spectrosopic signal in terms of a linear combination of dipole, quadrupole, and higher multipole contributions with known prefactors, allowing each multipole contribution to be recovered by statistical multivariate analysis of the experimental datasets \cite{Hu2008}.

An interesting case is that of an atom whose centre-of-mass is in a pure OAM state, such as in a Bose-Einstein condensate \cite{Andersen2006, Ramanathan2011}.  The OAM states of the atoms would then contribute a factor $e^{i(L-L')\phi_R}$ within the matrix element and we must integrate the factor $F_R$ with respect to the dynamical variable $\phi_R$. 
\begin{equation}
\int_0^{2\pi}e^{i(l+L-p-l'-L'+p')\phi_R}d\phi_R.
\end{equation}
This gives rise to a selection rule for OAM transfer involving the atomic centre of mass such that $\Delta p -\Delta L = \Delta l$.  $\Delta p$ then corresponds to the net OAM change induced in the atomic system and so we recover the selection rules derived in \cite{Lloyd2012, Lloyd2012b}.  We have indicated in \cite{Lloyd2013} that the work by \cite{Schattschneider2013} is not equipped to derive this selection rule as it misses the $\phi_R$ dependence in the matrix element.  One way to understand our result is to view the azimuthally delocalized state of the atom as interacting coherently with the vortex beam.  The cold atom gas has been subjected to electron beams \cite{Gericke2008}; our result suggests that CEVBS can be used as a test as to whether or not the atoms involved are in an OAM coherent state.

In summary, we have presented a new analysis of OAM transfer in inelastic atom-vortex interactions and derived the effective operator exhibiting quantised OAM transfer via multipolar excitations of the atom.  We demonstrated that the simplistic intepretation of dichroic spectroscopy based on equivalence of OAM change in the vortex beam with the corresponding change in the atomic electronic system is inapproriate without consideration of specific experimental situations because of the possible mode broadenning effect.  However, we have shown that the effect maybe minimized and nanoscale resolution chiral spectroscopy and spectral imaging are realisable either through optimization of the experimental set-up or through statistical multivariate analysis and that dichrotic spectroscopy can be equivalent to circular dichroism absorption and circularly polarized microscopy by photons and X-rays.  This will allow magnetic materials, chiral metamaterials or other chiral molecules to be studied in real space and in high resolution.  In addition, our results can be used to test for the coherence of cold atom systems through  the dependence of the spectroscopic selection rules on the nature of the centre-of-mass dynamics of the atoms.

\bibliography{ChiralBib}

\end{document}